# Molecular simulations of Perovskites CsXI$_3$ (X = Pb, Sn) Using Machine-Learning Interatomic Potentials

Atefe Ebrahimi[1], Franco Pellegrini[1], and Stefano De Gironcoli[1]

[1]Scuola Internazionale Superiore di Studi Avanzati (SISSA), Trieste, Italy
, aebrahim@sissa.it

**Abstract**

Cesium-based halide perovskites, such as CsPbI$_3$ and CsSnI$_3$, have emerged as exceptional candidates for next-generation photovoltaic and optoelectronic technologies, but their practical application is limited by temperature-dependent phase transitions and structural instabilities. Here, we develop machine-learning interatomic potentials (MLIPs) within the LATTE framework to simulate these materials with near experimental accuracy at a fraction of the computational cost compare to previous computaional study. Our Molecular dynamics simulations based on the trained MLIP, reproduce energies and forces across multiple phases, enabling largescale molecular dynamics simulations that capture cubic–tetragonal–orthorhombic transitions lattice parameters and octahedral tilting with unprecedented resolution. We find that Pb-based perovskites exhibit larger octahedral tilts and higher phase transition temperatures than Sn-based analogues, reflecting stronger bonding and enhanced structural stability, whereas Sn-based perovskites display reduced tilts and lower barriers, suggesting tunability through compositional or interface engineering. Beyond these systems, our work demonstrates that MLIPs can bridge first-principles accuracy with simulation efficiency, providing a robust framework for exploring phase stability, anharmonicity, and rational design in next-generation halide perovskites.

**Keywords:** CsSnI$_3$, machine-learned interatomic potentials, surface phase diagram, lead-free perovskites.

# 1  Introduction

Perovskites have emerged as a cornerstone of next-generation optoelectronic materials due to their tunable crystal structures, strong light absorption, and exceptional charge transport properties. These materials encompass diverse families, including metal halide perovskites (ABX$_3$), oxide perovskites (ABO$_3$), double perovskites (A$_2$BB'X$_6$), 2D layered perovskites (Ruddlesden–Popper and Dion–Jacobson phases), vacancy-ordered structures (A$_2$BX$_6$), and mixed-halide compositions [1–5].

However, despite their remarkable optoelectronic properties, the operational stability of perovskites remains a central challenge, primarily limited by temperature-induced lattice distortions and phase transitions that degrade device performance and lifetime.

Understanding and controlling these structural instabilities are therefore crucial for advancing perovskite-based technologies.

Among these, metal halide perovskites have garnered particular attention for photovoltaics and optoelectronics, combining excellent performance with low-temperature, solution-processable fabrication methods [6–10]. In contrast, oxide and other perovskites typically require high-temperature synthesis, limiting scalability and integration in devices. Table S1 shows representative examples of different perovskite families, their formulas, and main features.

Despite their potential, halide perovskites exhibit complex structural behavior at finite temperatures. Local lattice distortions and correlated octahedral tilting can form planar or three-dimensional structures, often deviating significantly from average crystallographic symmetry [11]. These dynamic fluctuations strongly influence electronic properties, charge transport, and device stability. [12–14].

By carefully selecting A-, B-, and X-site ions, key properties such as band gap, chargecarrier mobility, and environmental stability can be tuned, positioning these materials as promising candidates for efficient and durable energy technologies [15–18]. Among metal halide perovskites, cesium-based systems—particularly $CsPbI_3$ and $CsSnI_3$ offer fully inorganic compositions that enhance thermal and chemical stability, making them attractive for applications such as solar cells, light-emitting diodes, and photodetectors [19–21].

$CsPbI_3$ and $CsSnI_3$ represent particularly important cases. $CsPbI_3$ exhibits high charge-carrier mobility, strong light absorption, and favorable band gaps but suffers from instability in its high-symmetry cubic and tetragonal phases, which can irreversibly transform into non-perovskite phases under ambient conditions [22–24]. $CsSnI_3$, on the other hand, offers a lower-toxicity alternative to Pb-based compounds but also shows limited structural stability and lower phase-transition temperatures [25]. In both systems, phase behavior is governed by a combination of thermodynamic, kinetic, and defect-related factors [26–29]. Therefore, an accurate model that can provide details of atomistic insight into their structural dynamics is essential for both fundamental understanding and practical optimization.

Computational methods have emerged as an indispensable tool to explore molecular systems. While first-principles simulations, such as *ab initio* molecular dynamics, provide valuable insight into lattice dynamics and phase transitions, their computational cost restricts time and length scales. Conversely, empirical force fields lack the accuracy needed to capture the anharmonic and correlated lattice motions central to perovskite physics. To overcome these limitations, MLIPs have emerged as powerful tools capable of achieving near–DFT accuracy with orders-of-magnitude higher efficiency [30–36]. In halide perovskites, MLIPs have successfully reproduced local structural correlations, octahedral tilting, and phase transitions that are challenging to capture experimentally, provided the models are carefully trained and validated.

Several frameworks have been developed to implement MLIPs with varying levels of interpretability and computational cost. Among existing MLIP frameworks, such as SOAP [37], SNAP [38], and neural network potentials [39], the LATTE framework [40] provides key advantages: it offers simple and flexible descriptor construction, reduced computational overhead, and improved interpretability and transferability. These features make LATTE particularly suitable for investigating the structural dynamics of perovskites, where both high accuracy and scalability are critical.

In this work, we employ the LATTE descriptor in combination with atomic neural networks in the form of atomic multilayer perceptrons (MLPs), implemented within the PANNA package [41]. Using this framework, we systematically explore the structural dynamics and phase behavior of $CsPbI_3$ and $CsSnI_3$. The novelty of our study is threefold: (i) we directly compare Pb- and Sn-based perovskites, highlighting chemical trends in lattice dynamics and octahedral tilts; (ii) we perform large-scale, long-timescale molecular dynamics simulations, which are infeasible with conventional *ab initio* methods; and (iii) we provide a high-resolution analysis of octahedral tilt distributions and correlations, offering new atomistic insight into structural distortions and phase stability.

The remainder of this paper is organized as follows: Section 3.1 describes the training and validation of MLIPs for $CsPbI_3$ and $CsSnI_3$. Section 3.2 presents temperaturedependent pseudocubic lattice parameters and identifies cubic, tetragonal, and orthorhombic phase transitions. Section 3.3 reports detailed octahedral tilting analyses, revealing chemical trends between Pb- and Sn-based perovskites. Finally, Section 4 concludes with a summary of our findings, demonstrating that the proposed MLIP framework accurately captures phase transitions in both $CsPbI_3$ and $CsSnI_3$, in close agreement with experimental observations.

## 2 Computational Methods

### 2.1 Machine Learning Interatomic Potential Framework

The machine-learning interatomic potentials used in this study employ the LATTE descriptor [40], which encodes local atomic environments through Cartesian tensor contractions over spatially selected neighbors. This representation provides a compact yet accurate description of many-body interactions, and can be systematically extended through higher-order tensor contractions. The approach is computationally efficient and allows for training neural-network models to achieve near–DFT accuracy.

Prior to extending the methodology to novel systems, we carried out a validation of the MLIP by reproducing established results for cesium lead iodide ($CsPbI_3$). Specifically, we utilized the dataset reported by Baldwin *et al.* [42], which encompasses multiple crystallographic phases relevant to the phase transition behavior of $CsPbI_3$. In addition, the predicted octahedral tilting dynamics were benchmarked against the work of Eriksson and co-workers [43]. The close agreement with both studies, showing energy errors for $CsPbI_3$ and $CsPbBr_3$ of less than 10 meV/atom, demonstrates that the LATTE-based architecture provides a reliable and transferable description of halide perovskites.

Having established the reliability of the approach, we proceeded to construct a dataset for the lead-free perovskite $CsSnI_3$. Density functional theory (DFT) calculations were carried out using the Quantum ESPRESSO package [44]. Structural configurations were sampled across multiple crystallographic phases and over a broad temperature range, thereby ensuring adequate coverage of the relevant configurational space.

The final dataset comprises 214 training and 127 validation configurations for $CsPbI_3$, and 600 training and 200 validation configurations for $CsSnI_3$. Although the $CsSnI_3$ dataset is larger, the accuracy of our model remains excellent even with the smaller $CsPbI_3$ dataset, as demonstrated by energy and force parity plots that show close agreement with DFT reference values. Furthermore, molecular dynamics simulations and lattice predictions based on the trained models reproduce the expected phase behavior across temperatures, demonstrating their transferability to unseen conditions. Figure 1 compares the dataset compositions, illustrating the relative fractions of crystallographic phases represented for $CsPbI_3$ and $CsSnI_3$.

A complete list of hyperparameters and input files, including descriptor settings, network architecture, training variables, and raw datasets, is provided in our Zenodo repository (see Data Availability section).

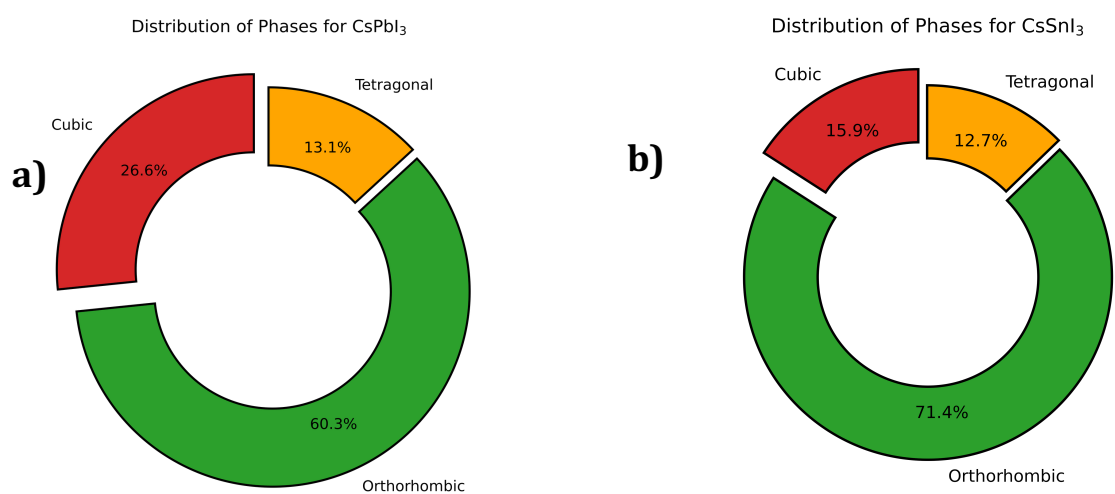

**Figure 1.** Comparison of dataset composition for a) $CsPbI_3$ and b) $CsSnI_3$.

## 2.2 DFT Calculations

The training dataset for $CsSnI_3$ was generated using DFT molecular dynamics in Quantum ESPRESSO to ensure diverse local atomic environments for robust MLIP training. We sampled 800 configurations covering multiple crystallographic phases and temperatures. Plane-wave basis sets were used with a kinetic energy cutoff of 60 Ry and a charge density cutoff of 425 Ry. Ultrasoft pseudopotentials (PBEsol) were employed for Cs, Sn, and I. Convergence thresholds were set to $10^{-7}$ Ry for total energy and $10^{-3}$ Ry/Bohr for forces. Brillouin zone sampling was performed on a 4×4×4 Monkhorst–Pack *k*-point grid. For $CsPbI_3$, we directly adopted the DFT training data from Baldwin *et al.* [42]. Input files for DFT calculations are available in our Zenodo repository (see Data Availability section).

## 2.3 MD Simulations

Molecular dynamics simulations were performed using the LAMMPS package [45] with the LATTE interatomic potential. The system comprised 4900 atoms, corresponding to a 7 × 7 × 5 repetition of the orthorhombic unit cell. Simulations were carried out in the $NpT$ ensemble with fully tri-axial cell fluctuations. Temperature was varied between 900 K and 50 K by performing both cooling (900 K → 50 K) and heating (50 K → 900 K) ramps, controlled via Nosé–Hoover dynamics with a damping constant of 0.1 ps. Pressure was maintained at 1 bar in all directions, with a damping constant of 1 ps. Each trajectory was run for 1 ns with a timestep of 1 fs. The chosen system size (4900 atoms) provides a reliable representation of bulk behavior and ensures that key structural and dynamical features are well converged, although extremely long-wavelength fluctuations may remain suppressed due to finite-size constraints. The resulting trajectories were subsequently analyzed to extract lattice parameters and octahedral tilt distributions, as described below.

## 2.4 Analysis of Lattice Parameters and Tilt Angles

To estimate lattice constants from molecular dynamics simulations, the systems were heated from 50 K to 900 K and subsequently cooled along the reverse path. The pseudocubic lattice parameter was monitored as a function of temperature, allowing identification of trends associated with structural phase transitions. The procedure used to compute the pseudocubic lattice parameter is described in detail in Section 3.2.

The octahedral tilt angles were analyzed following the procedure described by Larsen *et al.* [46]. First, M–X bonds (M = Pb or Sn, X = I) were identified to construct $MX_6$ octahedra. Each octahedron was then mapped onto an ideal cubic reference, yielding the rotation required to achieve the transformation. The rotation matrices were converted to Euler angles via quaternion representation using functionality implemented in ovito [47] and scipy [48]. Among the possible rotation conventions, we adopted the one yielding monotonically increasing tilt magnitudes, consistent with Glazer notation. This analysis was applied to each MD snapshot, enabling continuous tracking of tilt distributions as a function of temperature.

# 3 Results and Discussion

## 3.1 Machine Learning Potential Validation

One of the primary objectives of this study is to evaluate the reliability of machine learning–based interatomic potentials for molecular dynamics (MD) simulations, in comparison with conventional first-principles MD and empirical force-field approaches. As a necessary first step, an accurate potential must be constructed from a trained model. The machine learning (ML) model is first trained on total energy data and then refined using force information, ensuring a more faithful representation of the underlying potential energy surface. This two-step strategy improves the accuracy of force predictions, which is essential for generating realistic MD trajectories.

We demonstrate the accuracy of the model in predicting both energies and forces relative to DFT reference data. Once the predictive accuracy has been established, the model is employed to extract an interatomic potential, which is then used to carry out MD simulations.

The predictive quality of the model is quantified using the Root Mean Squared Error (RMSE), defined as

$$\text{RMSE} = \sqrt{\frac{1}{N}\sum_{i=1}^{N}\left(y_i^{\text{pred}} - y_i^{\text{true}}\right)^2}, \qquad (1)$$

where $y_i^{\text{pred}}$ represents the ML predictions, $y_i^{\text{true}}$ denotes the DFT reference values, and $N$ is the total number of data points.

The RMSE provides an absolute measure of the average prediction error, expressed in the same units as the target property. Because the squared errors are averaged, the RMSE is particularly sensitive to outliers and thus highlights deviations that may strongly influence the model's reliability.

Figure 2 presents the parity plot comparing ML-predicted forces with reference DFT forces. The data points lie closely along the $y = x$ line with $R^2 > 0.99$, indicating excellent agreement between ML and DFT. This suggests that the resulting interatomic potential is sufficiently accurate for use in MD simulations. Also for energy the reported $R^2$ correlation coefficients (see SI) further confirm the linear consistency between ML and DFT predictions.

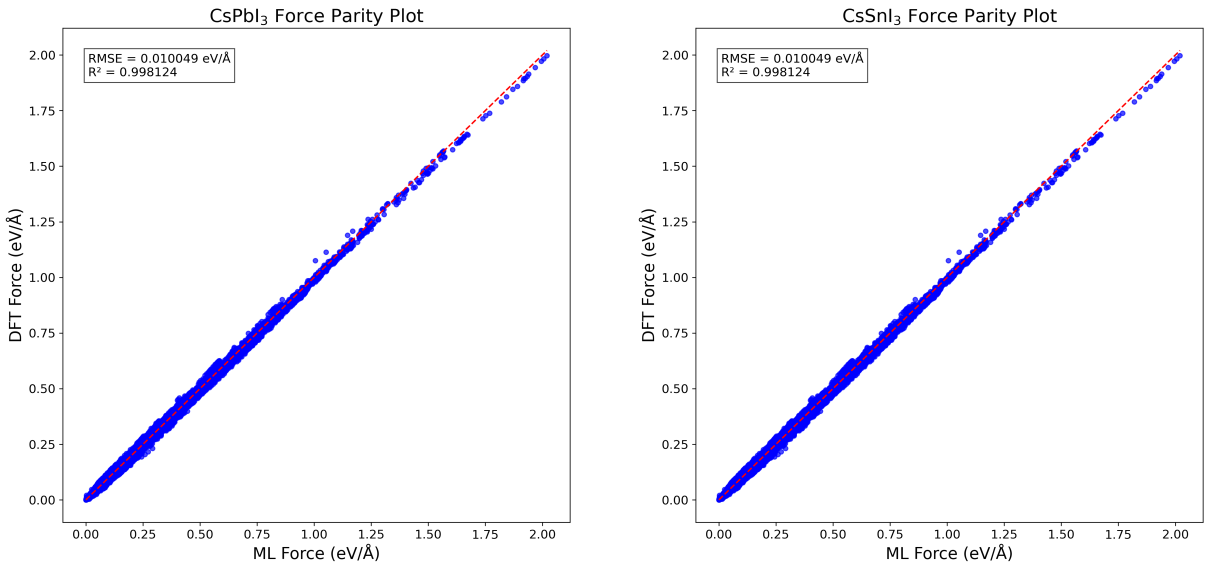

**Figure 2.** Parity plot of machine learning predicted forces compared with DFT reference forces for CsPbI$_3$ and CsSnI$_3$

Table 1 summarizes the RMSE values for energy (in eV/atom) and force (in eV/Å) predictions of the CsPbI$_3$ and CsSnI$_3$ surface models. Until now there are no papers that report the RMSE on force for bulk CsSnI$_3$ with MLIP study and same approach.

**Table 1.** Root-mean-square error (RMSE) for energy and force predictions of CsPbI3 and CsSnI3.

| Surface | Energy [eV/atom] | Force [eV/°A] |
|---|---|---|
| $CsPbI_3$(LATTE) | 0.0005 | 0.0261 |
| $CsSnI_3$(LATTE) | 0.0007 | 0.0100 |
| $CsPbI_3$[37] | 0.587 | 0.0310 |
| $CsSnI_3$ | - | - |

The LATTE model achieves RMSE values below 0.01 eV/atom and 0.03 eV/°A for energy and force predictions, respectively, which are comparable or superior to state-ofthe-art MLIPs for halide perovskites. This level of accuracy is sufficient to reproduce thermally induced phase transitions and local tilting behavior, as demonstrated in the following sections

## 3.2 Pseudocubic Lattice Parameter

The lattice constant, $a$, is one of the most fundamental structural parameters in perovskites, defining the size of the unit cell. Variations in the lattice constant arise from interactions between the constituent ions, influenced by their valence electrons and ionic radii. The lattice parameter strongly affects structural stability, octahedral tilting of the $BX_6$ framework, and electronic properties such as the band gap. Even small changes can significantly influence phase stability, transition temperatures, and optoelectronic performance. Monitoring the evolution of $a$, $b$, and $c$ with temperature during heating and cooling allows direct tracking of structural transformations in perovskites.

$CsPbI_3$ and $CsSnI_3$ exhibit four distinct structural phases, denoted as $\alpha$, $\beta$, $\gamma$, and $\delta$. For $CsPbI_3$, the first three phases correspond to the black perovskite phases with cubic ($\alpha$), tetragonal ($\beta$), and orthorhombic ($\gamma$) symmetry, whereas the $\delta$ phase is the yellow non-perovskite phase. Due to the relatively low Goldschmidt tolerance factor, the structural symmetry of $CsPbI_3$ decreases with temperature, leading to spontaneous conversion from black to yellow phases under ambient conditions. Stabilizing the black perovskite phases at room temperature is therefore essential for practical optoelectronic applications.

Similarly, $CsSnI_3$ exhibits high-temperature black perovskite phases and a low-temperature yellow non-perovskite phase. The cubic B-$\alpha$ phase at high temperature forms a threedimensional perovskite framework with $Sn^{2+}$ ions in ideal octahedral coordination, while Cs, Sn, and I atoms occupy regular lattice positions. Upon cooling, the cubic phase undergoes sequential symmetry reduction: the tetragonal B-$\beta$ phase arises due to octahedral tilting in the $ab$ plane, followed by the orthorhombic B-$\gamma$ phase with tilts along both apical and equatorial directions. The B-$\gamma$ phase is stable under inert conditions, while exposure to air or polar solvents converts it to the yellow Y phase, consisting of one-dimensional chains of edge-sharing $[Sn_2I_6]^{2-}$ octahedra separated by $Cs^+$ ions. Heating the yellow phase under inert conditions restores the cubic B-$\alpha$ phase [49].

Together, these observations illustrate that both $CsPbI_3$ and $CsSnI_3$ undergo temperaturedependent structural evolution from high-symmetry cubic phases to lower-symmetry tetragonal and orthorhombic perovskite phases, eventually converting to a yellow non-perovskite phase at ambient conditions. Understanding these transitions and stabilizing the black phases is critical for achieving robust optoelectronic performance.

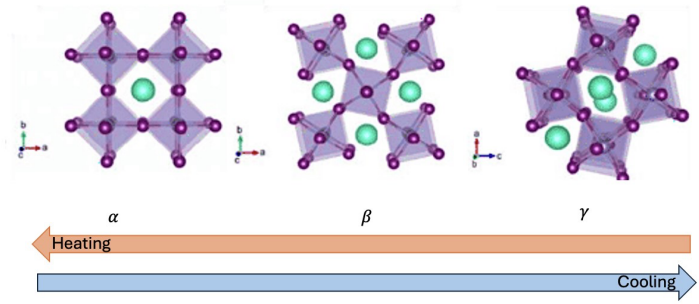

**Figure 3.** Schematic representation of temperature-dependent evolution of the lattice parameters in CsPbI$_3$ and CsSnI$_3$, illustrating the sequence of structural transitions. See Supplementary Video 1 for an animated visualization of lattice evolution.

Pb-based perovskites generally exhibit larger lattice parameters than Sn-based counterparts, correlating with increased octahedral tilting and enhanced structural stability.

To enable consistent comparisons across phases of different symmetry, we define a pseudocubic lattice parameter. For tetragonal and orthorhombic structures, the normalized parameters are expressed as

$$a_{norm} = \frac{a}{\sqrt{2}} \qquad b_{norm} = \frac{b}{\sqrt{2}} \qquad c_{norm} = \frac{c}{2} \qquad (2)$$

which highlight deviations from the cubic reference and facilitate visualization of lattice distortions with temperature [49–55].

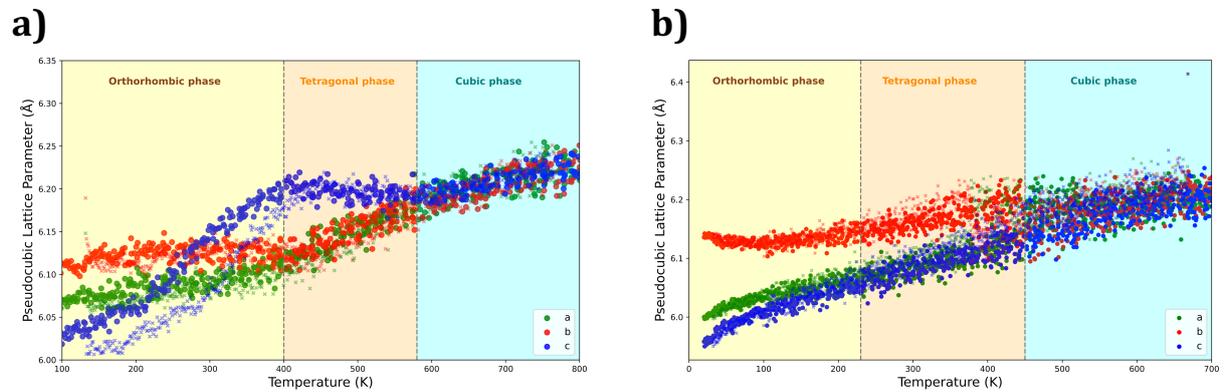

**Figure 4.** Pseudocubic lattice parameters versus temperature for (a) CsPbI$_3$ and (b) CsSnI$_3$. Circles denote heating trajectories, and crosses (×) denote cooling trajectories.

The temperature-dependent normalized lattice parameters were computed for both heating and cooling cycles. Figure 4 shows that the heating and cooling curves closely overlap for CsPbI$_3$, indicating minimal hysteresis. For CsPbI$_3$, three distinct phases are observed:

- **Orthorhombic phase:** At low temperatures (< 400 K), $a/= b/= c$. CsSnI$_3$ stabilizes in the orthorhombic phase below ∼ 230 K.

- **Tetragonal phase:** $CsPbI_3$ transitions to tetragonal symmetry between 400–580 K, with $a = b \neq c$. $CsSnI_3$ exhibits a comparable tetragonal phase between 230–450 K.

- **Cubic phase:** Above 580 K, $CsPbI_3$ adopts the cubic phase ($a = b = c$), persisting up to 800 K. $CsSnI_3$ reaches cubic symmetry above ∼ 450 K.

The same trends are observed during the cooling cycle, confirming the reversibility of the phase transitions. The phase transition temperatures and lattice parameters of $CsPbI_3$ closely reproduce previous computational predictions [42] and are in excellent agreement with experimental observations [56]. $CsSnI_3$ shows similar temperature-dependent behavior, with phase transitions occurring at lower temperatures (orthorhombic-to-tetragonal: ∼ 250–300 K; tetragonal-to-cubic: ∼ 400 K), consistent with experimental studies [57].

Overall, the MD simulations accurately capture the expected sequence of structural transitions, $\gamma \rightarrow \beta \rightarrow \alpha$, upon heating. At low temperatures, both compounds stabilize in the orthorhombic phase, whereas at high temperatures they adopt cubic symmetry. The normalized lattice parameter plots clearly illustrate these structural changes and associated phase transition temperatures.

## 3.3   Octahedral Tilting

Octahedral tilting of the corner-sharing $MX_6$ (M = Pb, Sn; X = I) units is a key structural feature in halide perovskites, critically influencing both phase stability and electronic properties. Abrupt changes in averaged tilt angles can serve as sensitive indicators of phase transitions. In this work, tilt angles were computed from MD simulations as described in Section 3.2 [46], using Euler angles to quantify rotations: $\theta = \phi$ for out-of-plane tilts along the $z$-axis, and $\psi$ for in-plane rotations within the $x$–$y$ plane.

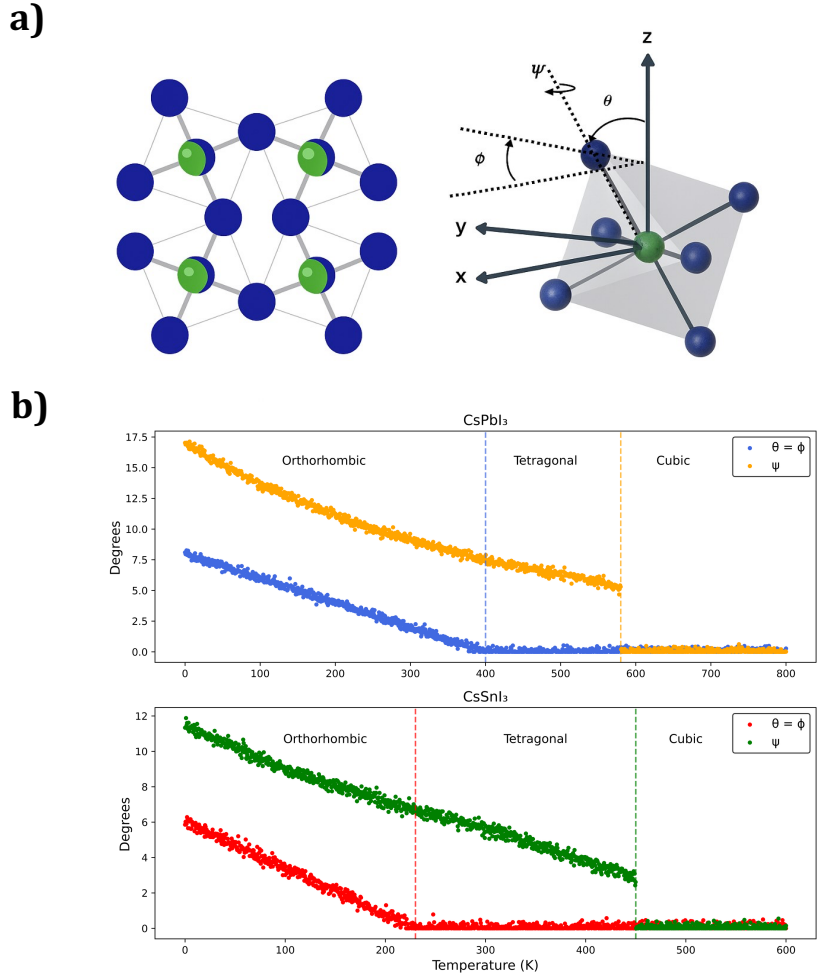

**Figure 5.** Pseudo-cubic lattice parameter vs. temperature for a) CsPbI$_3$ and b) CsSnI$_3$.

The temperature-dependent evolution of octahedral tilting in CsPbI$_3$ and CsSnI$_3$ follows the sequence of orthorhombic → tetragonal → cubic phases.

- **Orthorhombic phase (Pnma):** At low temperatures (T < 400 K for CsPbI$_3$, T < 230 K for CsSnI$_3$), all tilt angles are nonzero. In CsPbI$_3$, the out-of-plane tilt is $\theta = \phi = 6.2°$ and the in-plane rotation is $\psi = 11.9°$. In CsSnI$_3$, $\theta = \phi = 4.1°$ and $\psi = 7.2°$. The smaller tilts in CsSnI$_3$ are consistent with its lower phase transition temperatures.

- **Tetragonal phase (P4/mbm):** In CsPbI$_3$, tilting is restricted to the in-plane rotation $\psi = 9.8°$, while $\theta = \phi = 0$, over the 400–580 K temperature range. For CsSnI$_3$, the in-plane tilt is smaller, $\psi = 6.8°$, and out-of-plane tilts vanish, in the 230–450 K range.

- **Cubic phase (Pm$\bar{3}$m):** At high temperatures (T > 580 K for CsPbI$_3$, T > 450 K for CsSnI$_3$), all tilt angles reduce to zero, indicating perfectly aligned octahedra.

These results show that Sn-based perovskites consistently exhibit smaller octahedral distortions compared to Pb-based analogues. The reduced tilts correlate with lower orthorhombic-to-tetragonal and tetragonal-to-cubic transition temperatures, reflecting weaker structural rigidity. Furthermore, smaller tilts influence orbital overlap and lattice dynamics, which can affect the electronic and optoelectronic properties of these materials.

**Table 2.** Lattice parameters ($a$, $b$, $c$ in Å) and tilt angles ($\theta = \phi$, $\psi$ in degrees) for $CsPbI_3$ and $CsSnI_3$, with orthorhombic→tetragonal and tetragonal→cubic transition temperatures (K) from this work, experiments, and previous computations.

| Material | Phase | $a$ (Å) | $b$ (Å) | $c$ (Å) | $\theta = \phi$ (°) | $\psi$ (°) | Transition Temp. (K) |
|---|---|---|---|---|---|---|---|
| **$CsPbI_3$ (This Work)** | Orthorhombic | 6.070 | 6.113 | 6.018 | 6.2 | 11.9 | Ortho→Tetra: 400 |
| | Tetragonal | 6.111 | 6.120 | 6.205 | 0 | 9.8 | Tetra→Cubic: 580 |
| | Cubic | 6.214 | 6.234 | 6.216 | 0 | 0 | - |
| **$CsSnI_3$ (This Work)** | Orthorhombic | 6.030 | 6.133 | 6.000 | 4.1 | 7.2 | Ortho→Tetra: 230 |
| | Tetragonal | 6.160 | 6.148 | 6.144 | 0 | 6.8 | Tetra→Cubic: 450 |
| | Cubic | 6.414 | 6.414 | 6.414 | 0 | 0 | - |

Overall, the MD simulations reproduce the expected sequence of phase transitions, $\gamma \to \beta \to \alpha$, upon heating. The combination of tilt analysis and lattice parameters provides a consistent and quantitative description of structural evolution in $CsPbI_3$ and $CsSnI_3$, in excellent agreement with experimental and computational literature.

# 4    Conclusion

In this work, we have developed and validated MLIPs within the LATTE descriptor to investigate the structural dynamics and phase transitions of $CsPbI_3$ and $CsSnI_3$. By benchmarking against DFT datasets and experimental references, we demonstrated that the MLIPs faithfully reproduce energies, forces, and structural parameters, thereby reaching near-DFT accuracy at a fraction of the computational cost. Importantly, our parity analysis and root-mean-square error (RMSE) benchmarks establish a quantitative foundation for deploying these models in large-scale molecular dynamics simulations.

Using this framework, we systematically explored temperature-dependent phase transitions and octahedral tilting behavior in both lead- and tin-based perovskites. Our simulations reproduced the sequence of cubic–tetragonal–orthorhombic transitions, capturing the order and transition temperatures with close agreement to experimental and theoretical studies. A key outcome is the identification of clear chemical trends: $CsPbI_3$ exhibits larger octahedral tilts and higher transition temperatures compared to $CsSnI_3$, reflecting the stronger bonding and enhanced stability of Pb-based perovskites. Conversely, the lower tilting magnitudes and reduced transition barriers of $CsSnI_3$ highlight both its intrinsic instabilities and its potential tunability through compositional or interfacial engineering.

Beyond their immediate application to $CsPbI_3$ and $CsSnI_3$, our results underscore the broader utility of MLIPs for investigating halide perovskites. The ability to bridge firstprinciples accuracy with molecular dynamics efficiency enables unprecedented access to long timescale and large system simulations, providing insights into anharmonicity, local disorder, and phase competition that are otherwise difficult to capture. This methodological advance paves the way for predictive modeling of lead-free perovskites, where stability remains the primary bottleneck for practical photovoltaic applications.

Looking forward, the integration of MLIPs with advanced sampling techniques, defect chemistry, and device-level modeling will allow us to address open questions regarding degradation pathways, stabilization strategies, and composition–structure–property relationships. In this respect, the present study not only provides a validated potential for $CsPbI_3$ and $CsSnI_3$, but also establishes a robust framework for accelerating the discovery and optimization of next-generation, environmentally sustainable perovskite materials.

# 5    Data Availability

The machine learning potential, training and test datasets and example scripts on how to run the potential in LAMMPS movies of phase transition and extra information that mentiend in the text of paper are available at : doi:10.5281/zenodo.17201948

# Acknowledgements

This work was supported by SISSA and computational resources from CINECA and local HPC facilities.